\documentclass{IEEEtran}
\usepackage{graphicx}
\usepackage{amsmath}
\usepackage{amssymb}
\usepackage{float}
\begin{document}
\title{Finding a Zipf distribution and cascading propagation metric in utility line outage data }
\author{Ian Dobson, 
Iowa State University,  dobson@iastate.edu, August 2018}


\maketitle

\begin{abstract}
Observed  transmission line outage data is grouped into successive generations of events.
The empirical distribution of the number of generations in the cascades follows a Zipf distribution that implies the  increasing propagation as cascades progress. The slope of the Zipf distribution gives a System Event Propagation Slope Index (SEPSI). This new metric quantifies the cascade propagation, varies as expected, and determines the probabilities of small, medium, and large cascades.
\end{abstract}

\section{Introduction}

Long sequences of cascading outages occasionally cause large blackouts of power transmission systems. 
The cascade sequence starts with initial outages and is followed by propagating outages \cite{DobsonEPES17}. 
Cascading risk mitigation should address both the initiation and propagation of outages,
but there has been no single scalar metric that quantifies the propagation.
We find in historical utility data that the distribution of the number of cascading generations characterizes the propagation,
follows a Zipf distribution, and gives a new scalar metric of cascading propagation.

\section{Processing utility data into generations}

The transmission line outage data consists of 10942 automatic line outages recorded over 14 years by a North American utility~\cite{BPAwebsite, DobsonPS16}. 
The data includes the line outage start time to the nearest minute.
This data is standard and routinely collected by utilities worldwide, such as in the Transmission Availability Data System (TADS) in North America ~\cite{NERCreport14, BianPESGM14}.

The historical outage data is grouped into cascades and generations based on the outage start time using the simple method described in~\cite{DobsonPS12}. An outage occurring more than one hour after the preceding outage is assumed to start a new cascade, and within each cascade a series of outages less than one minute apart are grouped into the same generation. Thus each cascade consists of a series of generations of outage events, with each generation containing one or more line outages that occur closely spaced in time. For example,  outages caused by protection  within one minute are grouped together in the same generational event.
This processing produces 6687 cascades. The power system is usually resilient, so that most of these cascades are a  single generation of outages that does not propagate further.
\begin{figure}[h]
\centering
\includegraphics[width=\columnwidth]{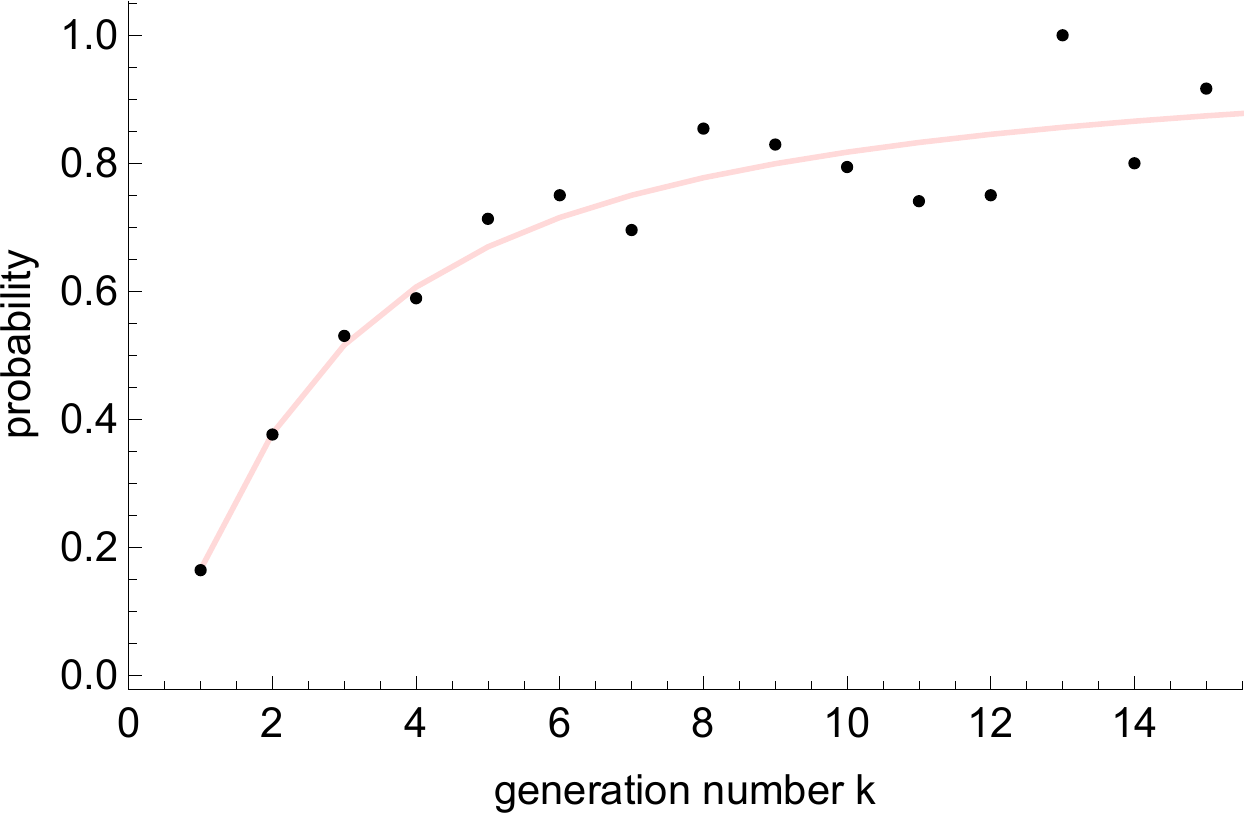}
\caption{Generation $k$ propagation $\rho_k$ as a function of $k$. Dots are utility data. Solid curve is calculated from the Zipf distribution solid line in Fig.~\ref{Gzipfdist}.}
\label{rhokdist}
\end{figure} 
\begin{figure}[h]
\centering
\includegraphics[width=\columnwidth]{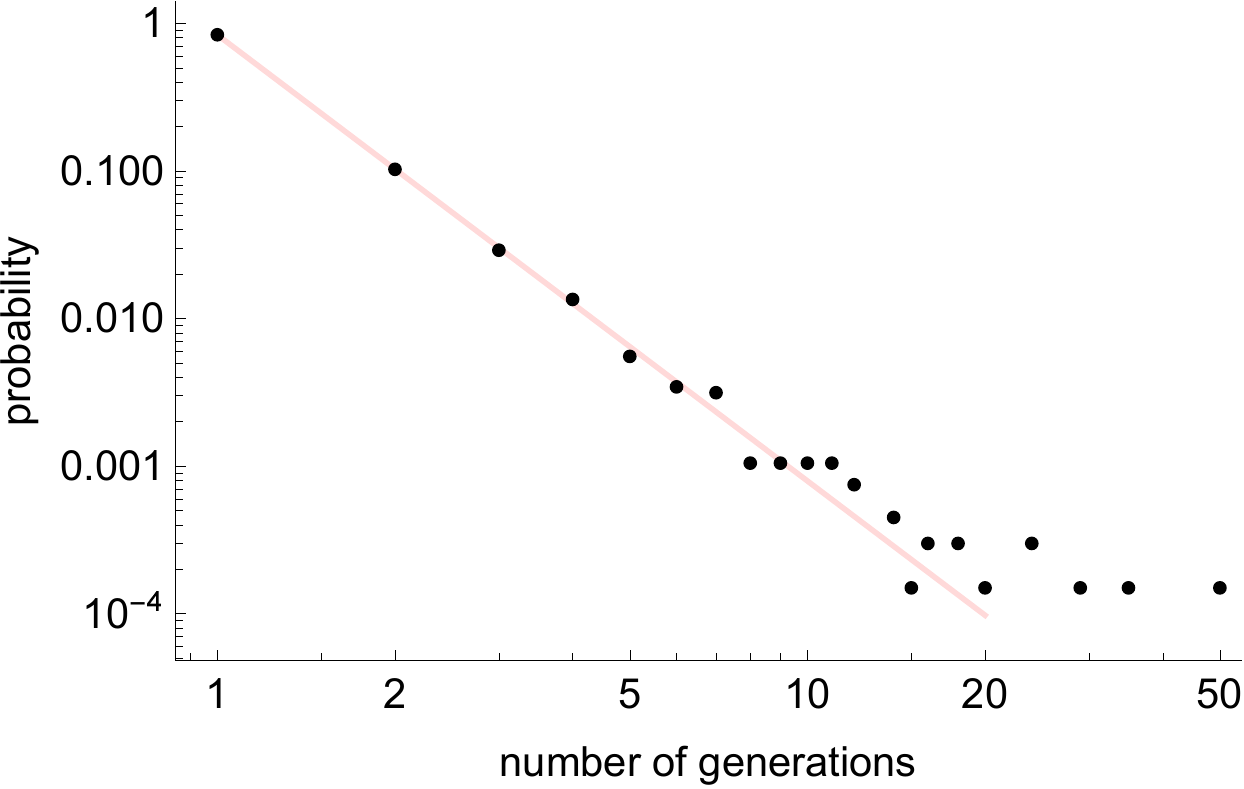}
\caption{Probability distribution of the number of generations on  log-log plot. Dots are utility data. Solid line indicates Zipf distribution best fitting  data.}
\label{Gzipfdist}
\end{figure}

\section{Propagation of generations and the Zipf distribution of the number of generations}

Instead of focussing on line outages as in previous work \cite{DobsonPS12}, 
here we analyze the propagation of generations of line outages.
Suppose a cascade has at least $k$ generations of outages and
let $\rho_k$ be the probability that generation $k$ propagates to produce a further generation  $k+1$.
The dots in Fig.~\ref{rhokdist} show how the propagation $\rho_k$ increases with $k$ in the data.

It is hard to characterize the increasing propagation in Fig.~\ref{rhokdist}  with a single number. 
Let us look at the propagation of generations in a different way.
In Fig.~\ref{Gzipfdist}
the dots show the empirical distribution of the number of generations $G$.
The distribution of $G$ is linear on this log-log plot and is a Zipf distribution (or zeta distribution) of the form
 \begin{align}
P[G=k]=\frac{1}{\zeta(s)}\frac{1}{k^{s}}, \quad k=1,2,3,...
\label{zipf}
\end{align}
where the slope of the line is $-s$ and $\zeta$ is the Riemann zeta function.
 Indeed, the maximum likelihood method of \cite{GoldsteinEPJB04} fits the data with the line of slope $-3.02$ shown in Fig.~\ref{Gzipfdist} and a Pearson $\chi^2$ test of goodness of fit shows consistency 
 with the Zipf distribution with p-value 0.88.
 
  Figs.~\ref{rhokdist} and  \ref{Gzipfdist} are different descriptions of the same propagation information, because  the hazard function of $G$ (the probability of the cascade stopping at generation $k$, 
  given that generation $k$ has been reached) is $1-\rho_k$.
  Indeed, the propagation $\rho_k$ implied by the fitted Zipf distribution of $G$ is indicated by the solid curve  in 
  Fig.~\ref{rhokdist}. Note how this solid curve interpolates the more erratic estimates from the sparse data 
  for the higher generations.
  
  \section{System Event Propagation Slope Index}
  
We propose using the negative of the slope of the line fitted to the distribution of the number of generations
 on the log-log plot as a cascading metric called the System Event Propagation Slope Index (SEPSI).
 For example, the slope of the fitted line in Fig.~\ref{Gzipfdist} is $-3.02$, so that  SEPSI\,$=$\,$3.02$.
A {\em lower} value of SEPSI indicates a shallower slope and 
 an {\em increased} probability of large cascades.
 
 The data for all automatic outages can be divided into two parts according to 
 whether the cascade occurs during a NOAA storm condition in the same weather zone,
or during the summer months June to September, or during peak load hours 3  to 8 pm \cite{DobsonHICSS18}.
Table~\ref{vary} shows  SEPSI calculated for each condition.
 As expected, SEPSI is smaller and cascading is more severe for the stressed cases, with the lowest value of SEPSI$=$$2.2$ achieved when there are storms.
 Moreover, as the condition varies, the  distribution of the number of generations remains linear on the log-log plot, as shown by the Pearson $\chi^2$ p-values in Table \ref{vary} for the fit of the Zipf distribution.
 
 We can use the Zipf distribution to deduce the probabilities 
 of small, medium, and large cascades from SEPSI. These probabilities are conditional on a cascade starting.
 Define a small cascade as 3 or less generations, a medium cascade as 4 to 9 generations, and a large cascade as 10 or more  generations
 (different cut-offs can be chosen).
 Then, substituting SEPSI for $s$ in (\ref{zipf}), we compute the probabilities $p_{\rm small}=\sum_{k=1}^3 P[G$$=$$k]$,  $p_{\rm medium}=\sum_{k=4}^9 P[G$$=$$k]$,  and $p_{\rm large}=1-p_{\rm medium}-p_{\rm large}$ as shown in Table \ref{vary}.  As expected, $p_{\rm large}$ varies with SEPSI by the largest factor. 
 
 To determine how many cascades  need to be observed for a given accuracy of SEPSI, we 
  bootstrapped samples from the Zipf distribution to find the approximate variation of the estimates of SEPSI.
   Suppose that there are $n$ cascades and SEPSI$=$3.0 so that  $p_{\rm large}$$=$$0.005$. 
 Then we find that the 95\% confidence interval of SEPSI is approximately $\pm 5/\sqrt{n}$.
 For example, with 95\%  confidence, 1000 cascades determine SEPSI within $\pm 0.16$ and $p_{\rm large}$ within a factor of 1.5.
 1000 cascades are needed for this accuracy because the larger cascades are rare.
 There are 782 cascades per year in our data, so that accumulating 1000 cascades takes 1.3 years.

\section{Conclusions}

The number of generations of events in a cascade is an indication of cascade size and severity.
We discover in utility line outage historical data that the distribution of the number of generations of events closely 
follows a Zipf distribution. 
This intriguing pattern in the data suggests using the negative of the slope of the Zipf distribution 
as a metric of cascading propagation, the System Event Propagation Slope Index or SEPSI.
For our data the overall SEPSI is 3.0 and this reduces to 2.2 in the presence of storms.
The probabilities of small, medium and large cascades can be computed from SEPSI.
Observing roughly 1000 cascades seems to determine  SEPSI and the probability of large cascades to a useful accuracy.

The results in this paper rely on line outages recorded by one utility.
But there is a clear and testable possibility of generalization to cascading events of a variety of components in other power systems and other infrastructures.
More general cascading events should be grouped into generations of events, and the empirical distribution 
of the number of the generations of cascading events should be obtained.
If the distribution of the number of generations can be satisfactorily approximated by a straight line on a log-log plot, then 
the cascading propagation and the probabilities of small, medium, and large cascades can be quantified with SEPSI.

\begin{table}[h]
	\caption{SEPSI varies with conditions and gives cascade probabilities}
	\label{vary}
	\centering
	\begin{tabular}{ cccccc }
		condition & SEPSI& p-value&$p_{\rm small}$&$p_{\rm medium}$&$p_{\rm large}$\\
		\hline
		all   & 3.0&0.88&0.967 & 0.029 & 0.005\\[0.7 mm]
	        storms  &2.2&0.29&0.877 & 0.086 & 0.037 \\
		no storms  &3.1&0.93&0.972 & 0.025 & 0.004 \\[0.7 mm]
		summer  & 2.9&0.98&0.961 & 0.033 & 0.006\\
		not summer  & 3.2&0.91&0.976 & 0.021 & 0.003\\[0.7 mm]
		peak hours  & 2.7&0.93&0.946 & 0.044 & 0.010\\
		non-peak hours  &3.1&0.94& 0.972 & 0.025 & 0.004\\
		\hline		
		\multicolumn{6}{c}{p-value indicates 
		fit of data to Zipf distribution}
	\end{tabular}
\end{table}


We gratefully acknowledge funding from NSF grants  1609080, 1735354
and thank BPA for making outage data  publicly available.
The analysis and conclusions are strictly the author's and not BPA's.



\end{document}